\documentclass{article}

\usepackage{booktabs}
\usepackage{array}
\usepackage{graphicx}
\usepackage{amsmath}
\usepackage{amsfonts}
\usepackage{textcomp}
\usepackage{multirow}
\usepackage{subfig}
\usepackage{multicol}
\usepackage{url}

\DeclareMathOperator*{\argmin}{arg\,min}

\begin{document}

\title{Bayesian design and analysis of external pilot trials for complex interventions}

\author{Duncan T. Wilson\textsuperscript{1} \and
	James M. S. Wason\textsuperscript{2,3} \and
	Julia Brown\textsuperscript{1} \and
	Amanda J. Farrin\textsuperscript{1} \and
	Rebecca E. A. Walwyn\textsuperscript{1}}
\date{1 - Leeds Institute of Clinical Trials Research, University of Leeds, Leeds, UK \\ 2 - Institute of Health and Society, Newcastle University, Newcastle, UK \\ 3 - MRC Biostatistics Unit, University of Cambridge, Cambridge, UK}

\maketitle

\begin{abstract}
External pilot trials of complex interventions are used to help determine if and how a confirmatory trial should be undertaken, providing estimates of parameters such as recruitment, retention and adherence rates. The decision to progress to the confirmatory trial is typically made by comparing these estimates to pre-specified thresholds known as progression criteria, although the statistical properties of such decision rules are rarely assessed. Such assessment is complicated by several methodological challenges, including the simultaneous evaluation of multiple endpoints, complex multi-level models, small sample sizes, and uncertainty in nuisance parameters. In response to these challenges, we describe a Bayesian approach to the design and analysis of external pilot trials. We show how progression decisions can be made by minimising the expected value of a loss function, defined over the whole parameter space to allow for preferences and trade-offs between multiple parameters to be articulated and used in the decision making process. The assessment of preferences is kept feasible by using a piecewise constant parametrisation of the loss function, the parameters of which are chosen at the design stage to lead to desirable operating characteristics. We describe a flexible, yet computationally intensive, nested Monte Carlo algorithm for estimating operating characteristics. The method is used to revisit the design of an external pilot trial of a complex intervention designed to increase the physical activity of care home residents.
\end{abstract}

\maketitle

\section{Introduction}\label{sec:introduction}

Complex interventions, defined as those comprised of several interacting components \cite{Craig2008}, can be challenging to evaluate in randomised controlled trials (RCTs) due to factors such as slow patient recruitment, poor levels of adherence to the intervention, and low completeness of follow-up data. To identify these problems prior to the main RCT we often conduct small trials \cite{Craig2008} known as pilots. These typically take the same form as the planned RCT but with a considerably lower sample size \cite{Eldridge2016}. If there is a seamless transition between the pilot and the main RCT, with all data being pooled and used in the final analysis, they are known as internal pilots. External pilots, in contrast, are carried out separately to the main RCT with a clear gap between the two trials. Pilot trials, which aim to inform the feasibility and optimal design of a subsequent definitive trial \cite{Lancaster2004}, are distinct from phase II trials, which focus instead on assessing potential efficacy and safety.

%\cite{Cooper2018} Are pilot trials useful for predicting randomisation and attrition rates in definitive studies: A review of publicly funded trials

%\cite{Morgan2018} Do feasibility studies contribute to, or avoid, waste in research?

%Whereas internal pilots are primarily used to inform sample size calculations, external pilots are used to address other feasibility aspects~\cite{Thabane2010}.

%\cite{Thabane2010}: ``internal pilot studies \ldots are primarily designed to inform sample size calculation for the main study \ldots [do] not usually address any other feasibility aspects.''

The data generated by an external pilot trial is used to help decide if the main RCT should go ahead, and if so, whether the intervention or the trial design should be adjusted to ensure success. In the UK, the National Institute for Health Research ask that these \emph{progression criteria} are pre-specified and included in the research plan \cite{NIHR2017}, and the recent CONSORT extension to randomised pilot trials requires their reporting \cite{Eldridge2016a}. A single pilot trial can collect data on several progression criteria, often focused on the aforementioned areas of recruitment, protocol adherence, and data collection~\cite{Avery2017}. Although they may take the form of single threshold values leading to binary stop/go decision rules, investigators are increasingly using two thresholds to accommodate an intermediate decision between stopping altogether and progressing straight to the main trial, which would allow progression but only after some adjustments have been made \cite{Eldridge2016a}. The need for appropriate progression criteria is clear when we consider the consequences of poor post-pilot progression decisions. If the criteria are too lax, there is a greater risk that the main trial will go ahead but found to be infeasible and thus a waste of resources; if the criteria are too strict, a promising intervention may be discarded under the mistaken belief that the main trial would be infeasible. Despite this, there is little published guidance about how they should be determined\cite{Avery2017, Hampson2017}.

In addition to pre-specifying progression criteria, another key design decision is the choice of pilot sample size. Conventional methods of sample size determination, which focus on ensuring the trial will have sufficient power to detect a target difference in the primary outcome, are rarely used since they would lead to a pilot sample size comparable with the main trial sample size. Several methods for pilot sample size determination instead aim to provide a sufficiently precise estimate of the variance in the primary outcome measure to inform the sample size of the main trial~\cite{Browne1995, Julious2005, Sim2012, Teare2014, Eldridge2015, Whitehead2015}. Others have suggested a simple rule of thumb for when the goal is to identify unforeseen problems~\cite{Viechtbauer2015}. While some have noted that the low sample size in pilots may lead to a considerable probability that a certain progression criterion will be met (or missed) due to random sampling variation \cite{Eldridge2015, Cooper2018}, and despite the consequences of making the wrong progression decision, the statistical properties of pilot decision rules are rarely used to inform the choice of sample size. This may be due to the methodological challenges commonly found in  pilot trials of complex interventions, including the simultaneous evaluation of multiple endpoints, complex multi-level models, small sample sizes, and prior uncertainty in nuisance parameters~\cite{Wilson2015}.

%\cite{Eldridge2016a}: ``This is a similar calculation to that used in estimating sample size needed for efficacy or effectiveness but allows for additional uncertainty in the resulting effect size estimate, thus effectively assessing potential effectiveness. If an objective is to assess potential effectiveness using a surrogate or interim outcome, investigators will need to use a standard sample size calculation to ensure there is adequate power. However, this type of objective is rare in pilot trials.''

%\cite{Cooper2018} Are pilot trials useful for predicting randomisation and attrition rates in definitive studies: A review of publicly funded trials

%\cite{Morgan2018} Do feasibility studies contribute to, or avoid, waste in research?

In this paper we will describe a method for designing and analysing external pilot trials which addresses these challenges. We take a Bayesian view, where progression decisions are made to minimise the expected value of a loss function. We propose a loss function with three parameters whose values can be determined either through direct elicitation of preferences or by considering the pilot trial operating characteristics they lead to. The operating characteristics we propose are all unconditional probabilities (with respect to a prior distribution) of making incorrect decisions, also known as assurances~\cite{OHagan2005}. Using assurances rather than the analogous frequentist error rates brings several benefits, including the ability to make use of existing knowledge whilst allowing for any uncertainty, and a more natural interpretation~\cite{Crisp2018}. As we will show, assurances are also useful when our preferences for different end-of-trial decisions are based on several attributes in a complex way that involves trading off some against others.

%\cite{Hee2016} Decision-theoretic designs for small trials and pilot studies: A review

The remainder of this paper is organised as follows. In Section~\ref{sec:methods} we describe the general framework for pilot design and analysis, some operating characteristics used for evaluation, and a routine for optimising the design. Two illustrative examples are then described in Sections~\ref{sec:TIGA} and~\ref{sec:REACH}. Finally, we discuss implications and limitations in Section~\ref{sec:discussion}.

\section{Methods}\label{sec:methods}

\subsection{Prior specification}

Consider a pilot trial which will produce data $x$ according to model $p(x | \theta)$. We decompose the parameters into $\theta = (\phi, \psi)$, where $\phi$ denotes the parameters of substantive interest and $\psi$ the nuisance parameters. We follow~\cite{Wang2002} and assume that two joint prior distributions of $\theta$ have been specified. First, the \emph{analysis} prior $p_A(\theta)$ is that which will be used when fitting the model once the pilot data is obtained. It has been argued that regulators are unlikely to accept the prior beliefs of the trial sponsor for analysis of the data \cite{OHagan2005, Walley2015}, and as such a weakly or non-informative prior should be used for $p_A(\theta)$ in order to ``let the data drive the inference'' \cite{Wang2002}. The choice of such a prior will be depend on the specific model being used, although methodological guidance for various specific cases such as logistic regression \cite{Gelman2008} and hierarchical models \cite{Spiegelhalter2001} is available. It should be emphasised, however, that the typically small sample size of a pilot trial can mean the effect of the analysis prior is non-negligible. As such, the analysis prior should provide a credible and justifiable representation of prior ignorance, avoiding extreme default choices which may place too much prior weight on infeasible regions of the parameter space.

The \emph{design} prior $p_D(\theta)$ will be used when evaluating the statistical performance of a proposed pilot trial design. It may be considered as purely hypothetical in the spirit of a `what-if' analysis \cite{Wang2002}, in which case several candidate design priors may be suggested and performance evaluated under each of these. Alternatively, and as we will assume in the remainder of this paper, $p_D(\theta)$ can be a completely subjective prior which fully expresses our knowledge and uncertainty in the parameters at the design stage. Although eliciting such a prior is potentially challenging, many examples describing successful practical applications of expert elicitation for clinical trial design are available~\cite{Walley2015, Crisp2018, Dallow2018}, as are tools for its conduct such as the Sheffield Elicitation Framework (SHELF)~\cite{OHagan2006a}.

\subsection{Analysis and progression decisions}\label{sec:analysis}

After observing the pilot data $x$, we must decide whether or not to progress to the main RCT. We consider three possible actions following the aforementioned `traffic light' system commonly used in pilot trials: 
\begin{itemize}
\item red - discard the intervention and stop all future development or evaluation; 
\item amber - proceed to the main RCT, but only after some modifications to the intervention, the planned trial design, or both; or
\item green - proceed immediately to the main RCT.
\end{itemize}

In what follows we will denote these decisions by $r, a$ and $g$ respectively. We assume that our preferences between the three possible decisions are influenced by $\phi$ but independent of $\psi$, formalising the separation of $\theta$ into substantive and nuisance components. We partition the substantive parameter space $\Phi$ into three subspaces $\Phi_I$, for $I=R,A,G$. Each subspace label corresponds to the decision we would make if we knew the true value of $\phi$. For example, if $\phi \in \Phi_R$ then the optimal decision is $r$(ed) - halt development and do not proceed to a definitive trial. We will henceforth refer to these three subsets as \emph{hypotheses}. Throughout, we will distinguish hypothesis $I$ from the corresponding optimal decision $i$ by using upper and lower case letters respectively.

When $\phi \in \Phi_{I}$ and we choose a decision $j \neq i$, there will be negative consequences. In particular, we may make three kinds of mistakes: proceed to an infeasible main RCT; discard a promising intervention; or make unnecessary adjustments to the intervention or trial design. We denote these errors as $E_1$, $E_2$, $E_3$ respectively. The occurrence of error $j$ will be denoted by $E_j = 1$, otherwise $E_j = 0$. An error's occurrence will be a function of the decision made $d$ and the true parameter value $\phi$, i.e. $E_j(d, \phi): \{r, a, g\} \times \Phi \rightarrow \{0,1\}$ for $j = 1,2,3$. We then use a loss function to express the preferences of the decision maker(s) on the space of possible events $E_1 \times E_2 \times E_3$ under uncertainty, defined as
$$
L(d, \phi) = c_1 E_1(d, \phi) + c_2 E_2(d, \phi) + c_3 E_3(d, \phi).
$$
Note that the additive form of the loss function implies that the our preferences for any one of the attributes $E_1, E_2, E_3$ are independent of the values taken by the others \cite{French2000}. 

To determine appropriate values of the parameters $c_1, c_2, c_3$ we first scale the loss function by setting $c_1 + c_2 + c_3 = 1$. Thus, a loss of 0 is obtained if no errors occur, and a loss of 1 is obtained if all errors occur (although note that this is not possible in this setting). We then follow the procedure described by French and Rios Insua (page 99) \cite{French2000}, eliciting some judgements from the decision maker(s) and using these to determine the values of $c_1, c_2, c_3$. One such judgement involves a simple gamble of obtaining the event $(E_1 = 0, E_2 = 0, E_3 = 0)$ with probability $1 - p_1$ and the event $(E_1 = 1, E_2 = 0, E_3 = 1)$ with probability $p_1$. The decision maker is asked to compare this gamble against an alternative of obtaining the event $(E_1 = 1, E_2 = 0, E_3 = 0)$ for certain, and to adjust the value of $p_1$ until they feel indifferent between the two options. Since this indifference implies the expected losses of the two options are equal, we will then have
$$
p_1 (c_1 +  c_3)  = c_1.
$$
Similarly, we can ask the decision maker(s) to consider a gamble between the event $(E_1 = 0, E_2 = 0, E_3 = 0)$ with probability $1 - p_2$ and the event $(E_1 = 1, E_2 = 1, E_3 = 0)$ with probability $p_2$, and compare this against the option of obtaining $(E_1 = 1, E_2 = 0, E_3 = 0)$ for certain. Again, by determining the value of $p_2$ which corresponds to indifference and thus equal expected loss, we deduce that
$$
p_2(c_1 + c_2) = c_1.
$$
This gives three equations which can be solved to obtain
$$
c_1 = \frac{-p_1 p_2}{p_1 p_2 - p_1 - p_2}, ~ c_2 = \frac{p_1 p_2 - p_1}{p_1 p_2 - p_1 - p_2} ~ c_3 = \frac{p_1 p_2 - p_2}{p_1 p_2 - p_1 - p_2}.
$$
Note that the two specific judgements suggested here are only two of many possible similar questions which could be posed to the decision maker(s). It is recommended that more indifferences are elicited in order to seek out any inconsistencies and further clarify their true preferences.

The loss function will then take values as given in Table~\ref{tab:loss}. For example, suppose we make a `green' decision under the `amber' hypothesis. The subsequent trial will be infeasible because the necessary adjustments will not have been made; but we have also discarded a promising intervention, since it would have been redeemed had the adjustments been made. The overall loss is therefore $c_{1} + c_{2}$.

\begin{table}
\caption{Losses associated with each decision under each hypothesis.}
\centering
\begin{tabular}{r r c c c}
\toprule
& & \multicolumn{3}{c}{Hypothesis} \\
& & $\phi \in \Phi_{R}$ & $\phi \in \Phi_{A}$ & $\phi \in \Phi_{G}$ \\
\midrule
\multirow{3}{*}{Decision} & $r$ & 0 & $c_{2}$ & $c_{2}$ \\
 & $a$ & $c_{1} + c_{3}$ & 0 & $c_{3}$ \\
 & $g$ & $c_{1}$ & $c_{1} + c_{2}$ & 0  \\
\bottomrule
\end{tabular}
\label{tab:loss}
\end{table}

%This allows us to define a function $a_{\mathbf{c}}(p_{R}, p_{G}): \mathcal{P} \rightarrow \mathcal{D}$ mapping the posterior probabilities $(p_{R}, p_{G}) \in \mathcal{P} = \{[0,1]^{2} \mid 0 \leq p_{R} + p_{G} \leq 1 \}$ (noting that $p_{A} = 1 - p_{R} - p_{G}$) to the optimal decision rule and parametrised by the cost vector $\mathbf{c} = (c_1, c_2, c_3)$.

Given a loss function with parameters $\mathbf{c} = (c_1, c_2, c_3)$ we follow the principle of maximising expected utility (or in our case, minimising the expected loss) when making a progression decision. We first use the pilot data in conjugation with the analysis prior $p_{A}(\theta)$ to obtain a posterior $p(\phi ~|~ x)$, and then choose the decision $i^{*}$ such that 
\begin{align}
i^{*} & = \argmin_{i \in \{r,a,g\}} \mathbb{E}_{\phi | x} [ L(i, \phi) ] \\
 & = \argmin_{i \in \{r,a,g\}} \int L(i, \phi) p(\phi | x) d\phi.
\end{align}
We can simplify this expression by noting that, given the piecewise constant nature of the loss function, the expected loss of each decision depends only on the posterior probabilities $p_{I} =  Pr[\phi \in \Phi_{I} ~|~ x]$ for $I = R, A, G$. We then have
\begin{align}\label{eqn:exp_loss}
\mathbb{E}_{\phi | x} [ L(r, \phi) ] & = p_{A}c_{3} + p_{G}c_{3}, \\
\mathbb{E}_{\phi | x} [ L(a, \phi) ] & = p_{R}c_{1} + p_{R}c_{2} + p_{G}c_{2}, \\
\mathbb{E}_{\phi | x} [ L(g, \phi) ] & = p_{R}c_{1} + p_{A}c_{1} + p_{A}c_{3}.
\end{align}
For some simple models which admit a conjugate analysis, the posterior probabilities $p_I$ can be obtained exactly. Otherwise, Monte Carlo estimates can be computed based on the samples from the joint posterior distribution generated by an MCMC analysis of the pilot data. Specifically, given $M$ samples $\phi^{(1)}, \phi^{(2)}, \ldots , \phi^{(M)} \sim p(\phi ~|~ x)$, 
\begin{equation}
p_I \approx \frac{1}{M} \sum_{k = 1}^{M}  \mathbb{I}(\phi^{(k)} \in \Phi_I),
\end{equation}
where $\mathbb{I}(.)$ is the indicator function.

\subsection{Operating characteristics}\label{sec:evaluation}

Defining a loss function and following the steps of the preceding section effectively prescribes a decision rule mapping the pilot data sample space $\mathcal{X}$ to the decision space $\{r, a, g\}$. To gain some insight at the design stage into the properties of this rule, we propose to calculate some trial operating characteristics. These take the form of unconditional probabilities of making an error when following the rule, calculated with respect to the design prior $p_D(\theta)$. We consider the following:
\begin{itemize}
\item $OC_1 = Pr[a ~\&~ \phi \in \Phi_R] + Pr[g ~\&~ \phi \in \Phi_R \cup \Phi_A]$ - probability of proceeding to an infeasible main RCT;
\item $OC_2 = Pr[r ~\&~ \phi \in \Phi_A \cup \Phi_G] + Pr[g ~\&~ \phi \in \Phi_A]$ - probability of discarding a promising intervention;
\item $OC_3 = Pr[a ~\&~ \phi \in \Phi_R \cup \Phi_G]$ - probability of making unnecessary adjustments to the intervention or the trial design.
\end{itemize}

These operating characteristics can be estimated using simulation. First, we draw $N$ samples $(\theta^{(1)}, x^{(1)}), (\theta^{(2)}, x^{(2)}), \ldots , (\theta^{(N)}, x^{(N)})$ from the joint distribution $p(\theta, x) = p(x | \theta)p_D(\theta)$. For each data set we then apply the analysis and decision making procedure described in Section~\ref{sec:analysis}, using some vector $\mathbf{c}$ to parametrise the loss function. This results in $N$ decisions $i^{(k)}$ which can be contrasted with the corresponding true parameter value $\theta^{(k)}$ and in which hypothesis it resides, noting if any of the three types of errors have been made. MC estimates of the operating characteristics can then be calculated as the proportion of occurrences of each type of error in the $N$ simulated cases. Assuming that $N$ is large, the unbiased MC estimate of an operating characteristic with true probability $p$ will be approximately normally distributed with variance $p(1-p)/N$.\footnote{Note that in the case of complex models which do no admit a conjugate analysis, the posterior probabilities obtained using an MCMC analysis will themselves be approximate and as such the optimal decision will be subject to error, which may increase the variance of the operating characteristic estimates. However, this issue can be sidestepped by assuming that, for each data set, the analysis that is simulated corresponds exactly to the analysis that would be carried out in practice. In particular, we assume that exactly $M$ posterior samples will be generated by the same MCMC algorithm, using the same seed in the random number generator.}

\subsection{Optimisation}\label{sec:optimisation}

Elicitation of the loss function parameters $\mathbf{c} = (c_1, c_2, c_3)$ in the manner described in Section \ref{sec:analysis} may be challenging, particularly when multiple decision-makers are involved~\cite{Keeney1976}. An alternative way to determine $\mathbf{c}$ is through examining the operating characteristics it leads to (for some fixed pilot design). As $\mathbf{c}$ is adjusted, the balance between the conflicting objectives of minimising each OC will change, and the task is then to find the $\mathbf{c}$ which returns the best balance from the perspective of the decision-maker. Formally, and thinking of operating characteristics as functions of $\mathbf{c}$, we wish to solve the multi-objective optimisation problem
\begin{equation}\label{eqn:opt}
\min_{\mathbf{c} \in \mathcal{C}} ~ \left( OC_{1}(\mathbf{c}),~ OC_{2}(\mathbf{c}),~ OC_{3}(\mathbf{c}) \right)
\end{equation}
where $\mathcal{C} = \{c_{1}, c_{2} \in [0,1] ~|~ c_{1} + c_{2} \leq 1\}$. 

Since the three objectives are in conflict there will be no single solution which simultaneously minimises each one. We would instead like to find a set $\mathcal{C}^* = \{ \mathbf{c}^{(1)}, \mathbf{c}^{(2)}, \ldots, \mathbf{c}^{(K)} \}$ such that each member provides a different balance between minimising the three operating characteristics. If there exist $\mathbf{c}, \mathbf{c}' \in \mathcal{C}^*$ such that $OC_i(\mathbf{c}') \leq OC_i(\mathbf{c})$ for all $i \in \{1, 2, 3\}$ and $OC_i(\mathbf{c}') < OC_i(\mathbf{c})$ for some $i \in \{1, 2, 3\}$, we say that $\mathbf{c}'$ dominates $\mathbf{c}$. In this case, because $\mathbf{c}$ leads to worse (or at least no better) values of all three operating characteristics when compared to $\mathbf{c}'$, we have no reason to include it in our set $\mathcal{C}^*$. Because the search space $\mathcal{C}$ has only two dimensions, problem (\ref{eqn:opt}) can be approximately solved by generating a uniform random sample of $\mathbf{c}$'s and estimating the operating characteristics for each. Any parameters which are dominated in this set can then be discarded, and the operating characteristics of those which remain can be illustrated graphically. The decision maker(s) can then view the range of available options, all providing different trade-offs amongst the three operating characteristics, and choose from amongst them. 

To solve the problem in a timely manner we must be able to estimate operating characteristics quickly. Noting from equation (\ref{eqn:exp_loss}) that the expected loss of each decision depends only on $\mathbf{c}$ and the posterior probabilities $p_R, p_A$ and $p_G$, we first generate $N$ samples of these posterior probabilities and then use this same set of samples for every evaluation. This approach not only ensures that optimisation is computationally feasible, but also means that differences in operating characteristics are entirely due to differences in costs, as opposed to differences in the random posterior probability samples.

\section{Illustrative example - Child psychotherapy (TIGA-CUB)}\label{sec:TIGA}

TIGA-CUB (Trial on Improving Inter-Generational Attachment for Children Undergoing Behaviour problems) was a two-arm, individually-randomised, controlled pilot trial informing the feasibility and design of a confirmatory RCT comparing Child Psychotherapy (CP) to Treatment as Usual (TaU), for children with treatment resistant conduct disorders. The trial aimed to recruit $60$ primary carer-child dyads, to be randomised equally to each arm. This sample size was chosen to give desired levels of precision in the estimates of the common standard deviation of the primary outcome, the follow-up rate, and the adherence rate. Here, we focus on the latter two parameters and consider how our proposed method could have informed the design of TIGA-CUB.

We model the number of participants successfully followed-up (denoted $f$) using a binomial distribution with parameter $p_f$, and similarly the number successfully adhering to the intervention (denoted $a$) with a binomial distribution with parameter $p_a$. For a fixed pilot trial per-arm sample size $n$, the parameters of the model are $\phi = (p_f, p_a)$, with no nuisance parameters. Further assuming that the numbers followed-up and adhering are independent, the likelihood is then 
$$
p(f, a | p_f, p_a) = \left[{2n \choose f}p_f^{f}(1-p_f)^{2n-f}\right] \times \left[{n \choose a}p_a^{a}(1-p_a)^{n-a}\right].
$$

At the design stage, the follow-up rate $p_f$ was thought to be somewhere in the range 62\% to 92\%, while the adherence rate $p_a$ was thought to lie between 40\% and 95\%. We reflect these ranges of uncertainty in our design priors by using beta distributions $p_f \sim Beta(40, 10)$ (thus giving a prior mean of 0.8), and $p_a \sim Beta(11.2, 4.8)$ (giving a prior mean of 0.7). We assume that a uniform `non-informative' prior $Beta(1,1)$ will be used for each parameter in the analysis.

%Quantitative progression criteria were not pre-specified in TIGA-CUB. 

TIGA-CUB's progression criteria included only simple stop/go thresholds, with no intermediate `amber' decisions. As such, in this example we partition the parameter space into two hypotheses, $\Phi_G$ and $\Phi_R$. For the purposes of illustration we define the hypothesis $\Phi_G$ as the subset of the parameter space where $p_f >= 0.8$ and $p_a >= 0.7$, hypothesis $\Phi_R$ being its complement. Thus, in this example we do not consider there to be a trade-off between the two parameters of interest. For the main trial to be feasible, both must be above their respective thresholds. The prior distributions on parameters $p_f$ and $p_a$ imply an \emph{a priori} probability of 0.28 that $\phi \in \Phi_G$, i.e. that both follow-up and adherence are sufficiently high.

In this special case, the loss function is 
$$
L(d, \phi) = c_1 E_1(d, \phi) + c_2 E_2(d, \phi)
$$
and the expected losses of decisions $g$ and $r$ will be $\mathbb{E}_{\phi | x}[L(g, \phi)] = c_1 p_R$ and $\mathbb{E}_{\phi | x}[L(r, \phi)] = c_2 p_G $, where $p_R + p_G = 1$ and $c_1 + c_2 = 1$. Decision $g$ is therefore optimal whenever $p_G > c_1$. The posterior probability $p_G$ can be easily calculated given the pilot data due to the beta prior distributions being conjugate. Specifically, given a total sample size $n$ and observing $x_f$ participants with follow-up and $x_a$ participants with adherence, the posterior probability $Pr[\phi \in \Phi_G ~|~ x]$ is given by
\begin{equation}
p_G = [1 - F(0.8; 1+x_f, 1+n-x_f)] \times [1 - F(0.7; 1+x_a, 1 + n/2 - x_a)],
\end{equation}
where $F(y; \alpha, \beta)$ denotes the cumulative probability function of the beta distribution with parameters $\alpha, \beta$.

At the design stage we can calculate the probability of an infeasible trial ($OC_1$),
\begin{align}
Pr[g, \phi \in \Phi_R] &= \int_{\Phi_R} Pr[g ~|~ \phi] p(\phi) d\phi \\
 &= \int_{\Phi_R} \left( \sum_{x_f = 0}^{n} \left[ \sum_{x_a = 0}^{n/2}  \mathbb{I}(p_G < c_1 ~|~ x_f, x_a, n) p(x_a ~|~ \phi) \right]p(x_f ~|~ \phi) \right)p(\phi) d\phi,
\end{align}
and similarly for the probability of discarding a promising intervention. As these calculations can be computationally expensive for moderate $n$ due to the nested summation term, we use Monte Carlo approximations as described in Section~\ref{sec:methods}. 

\begin{figure}
\centering
\includegraphics[scale=0.8]{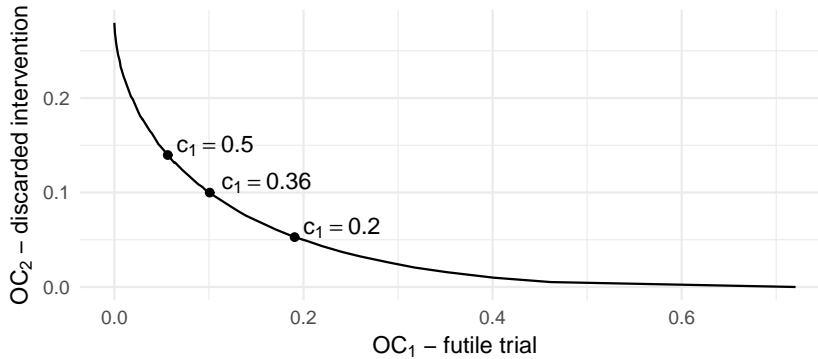}
\caption{Probabilities of an infeasible main trial ($OC_1$) and of discarding a promising intervention ($OC_2$) for a range of loss parameters $c_1$ when sample size is fixed at $n=30$.}
\label{fig:tiga_n60}
\end{figure}

Keeping the sample size fixed at $n = 30$ per arm, we estimated the operating characteristics using a range of cost parameters values $c_1 = 0, 0.02, 0.04, \ldots , 1$ using $M = 10^6$ Monte Carlo samples. The results are plotted in Figure~\ref{fig:tiga_n60}, with some specific values of $c_1$ highlighted. The decision-maker can decide which point on the operating characteristic curve best reflects their own priorities in terms of the two types of error. For example, if the consequences of running an infeasible main RCT are considered less important than those of needlessly discarding a potentially effective intervention, the decision-maker may choose to set $c_1 = 0.2$ and would obtain $OC_1 = 0.19, OC_2 = 0.05$.

%\begin{table}
%\centering
%\caption{Estimated operating characteristics of the TIGA-CUB trial with sample size $n=30$ for the three loss parameter values highlighted in Figure~\ref{fig:tiga_n60}. Figures are rounded to 2 decimal places, and all standard errors are $< 0.0005$.}
%\input{tables/tiga_points.txt}
%\label{tab:tiga_points}
%\end{table}

\begin{figure}
%\begin{multicols}{3}
\centering
\includegraphics[height=4cm, trim={0 0 2.5cm 0},clip]{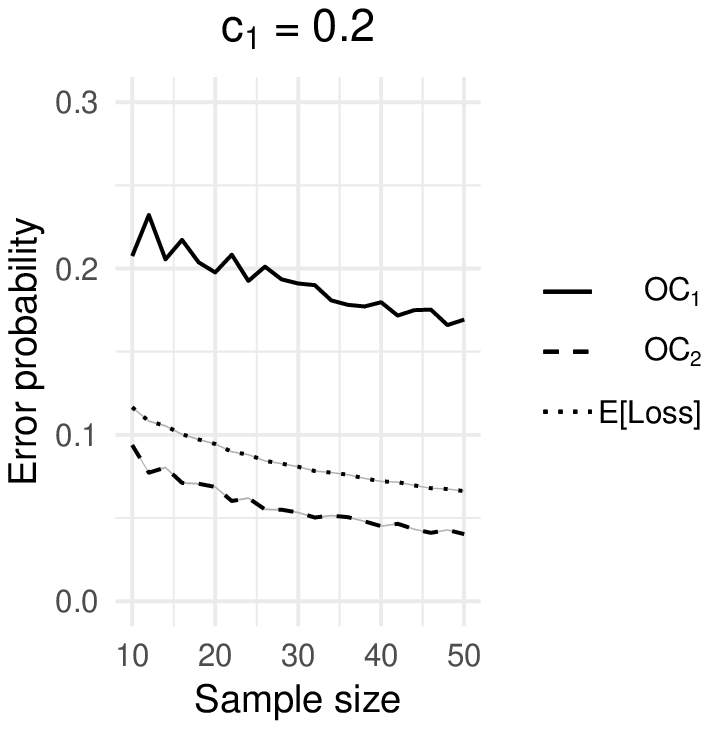}\hspace{0.5cm}
\includegraphics[height=4cm, trim={1.1cm 0 2.5cm 0},clip]{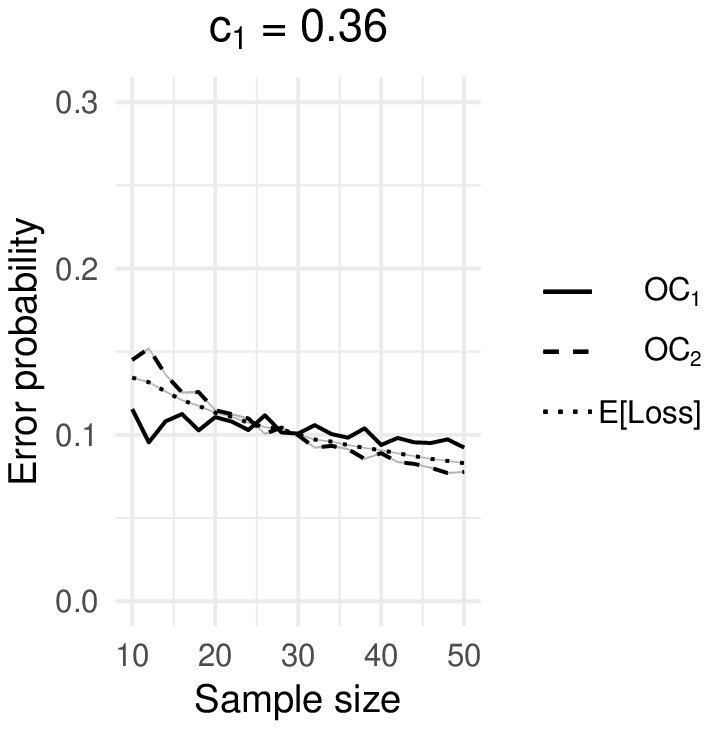}\hspace{0.5cm}
\includegraphics[height=4cm, trim={1.1cm 0 0 0},clip]{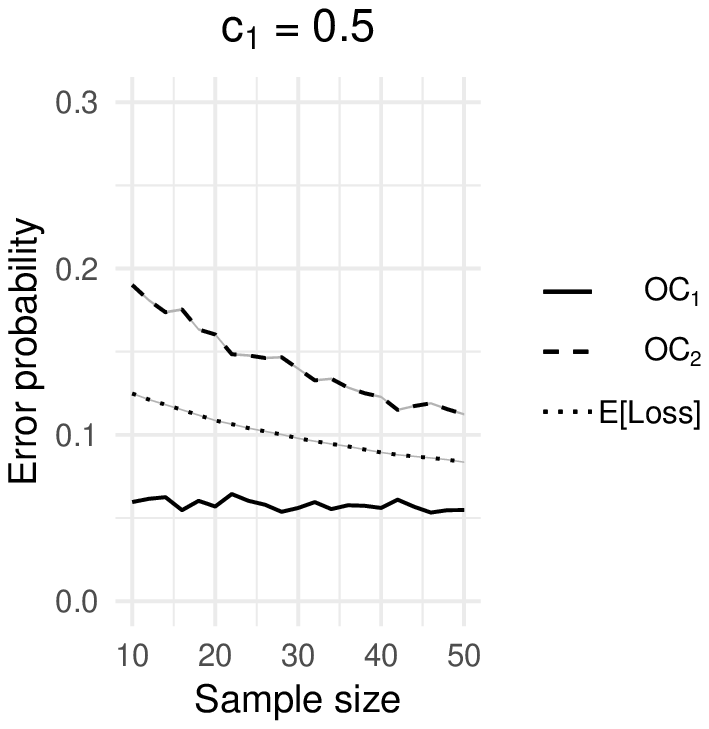}
%\end{multicols}
\caption{Probabilities of an infeasible main trial ($OC_1$) and of discarding a promising intervention ($OC_2$) for a range of per-arm sample sizes and different values of the loss parameter $c_1$.}
\label{fig:tiga_ocs}
\end{figure}

To examine the effect of adjusting the sample size, we evaluated the operating characteristics obtained for $n = 10, 12, 14, \ldots , 50$ per arm whilst setting $c_1 = 0.2, 0.36, 0.5$. The results are shown in Figure~\ref{fig:tiga_ocs}. Each line includes a shaded area denoting the 95\% Monte Carlo error intervals, although these are so small as to be illegible given the high number ($M = 10^6$) of MC samples used for each calculation. Although operating characteristics generally improve as the sample size is increased, we see that for $c_1 = 0.36$ and 0.5 the probability of an infeasible main trial, $OC_1$, remains flat whilst $OC_2$ has a downward trend. As we would expect, the  the expected loss reduces smoothly as $n$ increases in all cases. In contrast, there is some variability beyond that explained by MC error in the OCs. This can be explained by the discrete nature of simulated adherence and follow-up data. Our results show that, for the design priors and hypotheses used in this example, the chosen sample size in TIGA-CUB of $n=30$ can provide error rates broadly in line with conventional type I and II error rates under the usual hypothesis testing framework.

\section{Illustrative example - Physical activity in care homes (REACH)}\label{sec:REACH}

The REACH (Research Exploring Physical Activity in Care Homes) trial aimed to inform the feasibility and design of a future definitive RCT assessing a complex intervention designed to increase the physical activity of care home residents~\cite{Forster2017}. The trial was cluster randomised at the care home level, with twelve care homes in total randomised equally between treatment as usual (TaU) and the intervention plus TaU.

Data on several feasibility outcomes were collected. Here, we focus on four: recruitment (measured in terms of the average number of residents in each care home who participate in the trial, or average cluster size); adherence (a binary indicator at the care home level indicating if the intervention was fully implemented); data completion (a binary indicator for each resident of successful follow-up at the planned primary outcome time of 12 months); and potential efficacy (a continuous measure of physical activity at the resident level). Progression criteria using the traffic light system were pre-specified for all of these outcomes except potential efficacy, as detailed in Table~\ref{tab:pcs}

\begin{table}
\caption{Pre-specified progression criteria used in the original REACH design.}
\centering
\begin{tabular}{r l l l}
\toprule
Outcome & Red & Amber & Green \\
\midrule
Recruitment (avg. per care home) & Less than 8 & Between 8 and 10 & At least 10 \\
Adherence & Less than 50\% & Between 50 and 75\% & At least 75\% \\
Follow-up & Less than 65\% & Between 65 and 75\% & At least 75\% \\
\bottomrule
\end{tabular}
\label{tab:pcs}
\end{table}

Denoting the size of the $j$th cluster by $m_j$, we assume that cluster sizes are normally distributed, $m_j \sim N(\mu_c, \sigma^2), j = 1, \ldots , 2k$. We further assume that the probability of a participant being followed-up is constant across clusters and arms, and that the total number follows a binomial distribution $f \sim Bin(\sum_{j=1}^{2k} m_j, p_f)$. The number of care homes which successfully adhere to the intervention is assumed to binomially distributed, $a \sim Bin(k, p_a)$.

The continuous measure of physical activity is expected to be correlated within care homes. We model this using a random intercept, where the outcome $y_{ij}$ of resident $i$ in care home $j$ is
\begin{equation}
y_{ij} = X_{j} \times Y_{j} \times \mu + u_{j} + \varepsilon_{i}.
\end{equation}
Here, $X_{j}$ is a binary indicator of care home $j$ being randomised to the intervention arm, $Y_{j}$ is a binary indicator of care home $j$ successfully adhering to the intervention, $\mu$ is the mean treatment effect, $u_{j} \sim \mathcal{N}(0, \sigma_{B}^{2})$ is the random effect for care home $j$, and $\varepsilon_{i} \sim \mathcal{N}(0, \sigma_{W}^{2})$ is the residual for resident $i$. We parametrise the model using the intracluster correlation coefficient, $\rho = \sigma_{B}^{2} / (\sigma_{B}^{2} + \sigma_{W}^{2})$.

The parameters describing average cluster size, follow-up and adherence rates, and mean treatment effect are of substantive interest when making progression decisions, giving $\phi = (\mu_c, p_f, p_a, \mu)$. The remainder are nuisance parameters, $\psi = (\sigma^2, \rho, \sigma_W^2)$.

\subsection{Model specification}

To begin specifying a model for the REACH trial, we first note that the four substantive parameters can be divided into two pairs. Firstly, mean cluster size and follow-up rate relate to the amount of information which a confirmatory trial will gather. Secondly, potential efficacy and adherence relate to the effectiveness of the intervention, where effectiveness is thought of as the effect which will be obtained in practice when the effect of non-adherence is accounted for. We expect that a degree of trade-off between adherence and potential efficacy will be acceptable, with a decrease in one being compensated by an increase in the other. Likewise, low mean cluster size could be compensated to some extent by higher follow-up rate, and vice versa. 

While there may be trade-offs within these pairs of parameters, we do not expect trade-offs between them. A trial with no effectiveness will be futile regardless of the amount of information collected, and so should not be conducted. Similarly, a confirmatory trial should not be conducted if it is highly unlikely to produce enough information for the research question to be adequately answered. We therefore consider the sub-spaces of $\Phi$ formed by these parameter pairs, partition these into hypotheses, and combine these together. Constructing hypotheses in these two-dimensional spaces is cognitively simpler than working in the original four dimensional space, not least because they can be easily illustrated graphically.

Formally, let $\Phi^i$ be the sub-space of mean cluster size and follow-up rate, and $\Phi^e$ be that of adherence and potential efficacy. Having specified hypotheses $\Phi^i_I, \Phi^e_I$ for $I = R,A,G$, we then have 
\begin{align}\label{eqn:comb_hyp}
\phi \in \begin{cases}
               \Phi_R \text{ if }  \phi^i \in \Phi^i_R \text{ or } \phi^e \in \Phi^e_R \\
               \Phi_G \text{ if }  \phi^i \in \Phi^i_G \text{ and } \phi^e \in \Phi^e_G \\
               \Phi_A \text{ otherwise}. \\
            \end{cases}
\end{align}

\subsubsection{Follow-up and cluster size}

Recall that cluster sizes are assumed to be normally distributed with mean $\mu_c$ and variance $\sigma^2)$. A normal-inverse-gamma prior 
\begin{equation}
\sigma^{2} \sim \Gamma^{-1} (\alpha_{0}, \beta_{0}), ~ \mu_{c} \sim N(\mu_{0}, \sigma^{2}/\nu_{0})
\end{equation}
is placed on the mean and variance to allow for prior uncertainty in both parameters. It was anticipated that an average of 8 - 12 residents would be recruited in each care home. To reflect this prior belief we set the hyper-parameters to $\mu_{0} = 10, \nu_{0} = 6, \alpha_{0} = 20, \beta_{0} = 39$, giving a prior cluster size of 10 with mean variance 2.05.

For the probability of successful follow-up, $p_f$, we take a Beta distribution with hyper-parameters $\alpha_{0} = 22.4, \beta_{0} = 9.6$ as the prior. This gives a prior with a mean of 0.7 and a standard deviation of 0.08.

To partition the parameter space into hypotheses, we first consider the case where follow-up is perfect, i.e. $p_{f} = 1$. Conditional on this, we reason that a mean cluster size of below 5 should lead to a red decision (stop development), whereas a size of above 7 should lead to a green decision (proceed to the main trial). As the probability of successful follow-up decreases, we suppose that this can be compensated by an increase in mean cluster size. We assume the nature of this trade-off is linear and decide that if $p_{f}$ were reduced to 0.8, we would want to have a mean cluster size of at least 8 to consider decisions $a$ or $g$.  We further decide that a follow-up rate of less than $p_{f} = 0.6$ would be critically low, regardless of the mean cluster size, and should always lead to decision $r$. Similarly, a follow-up rate of $0.6 \leq p_{f} < 0.66$ should lead to modification of the intervention or trial design. Together, these conditions lead to the following partitioning of the parameter space:
\begin{equation}
  (p_{f}, \mu_{c}) \in \begin{cases}
               \Phi^i_R \text{ if } p_{f} < 0.6 \text{ or } 20-15p_{f} > \mu_{c} \\
               \Phi^i_G \text{ if } p_{f} > 0.66 \text{ and } 22-15p_{f} < \mu_{c} \\
               \Phi^i_A \text{ otherwise.}
            \end{cases}
\end{equation}

\begin{figure}
   \centering
   \subfloat[][]{\includegraphics[height=4cm, trim={0 0 2.8cm 0},clip]{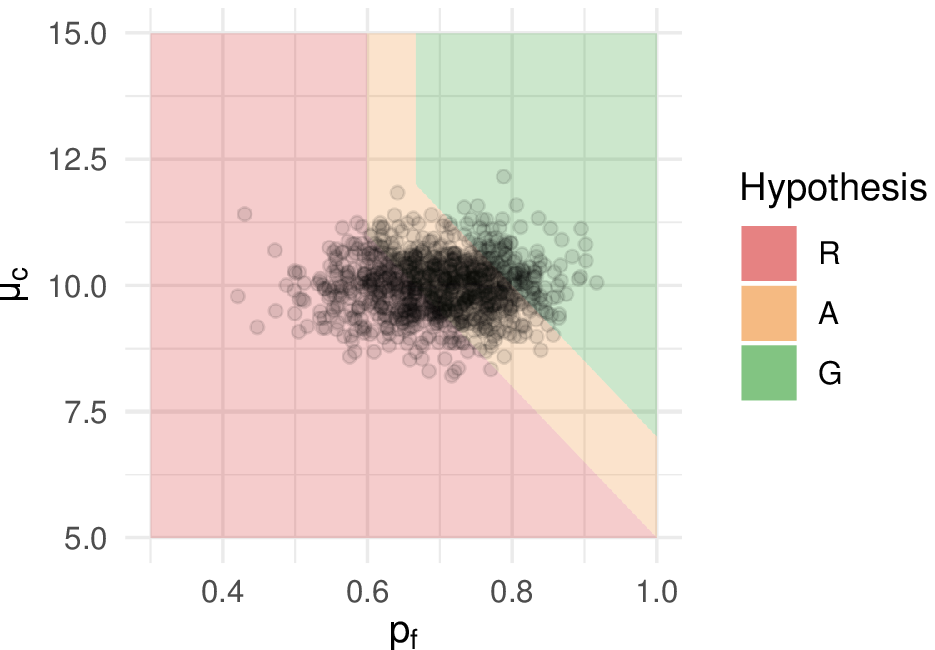}}\quad
   \subfloat[][]{\includegraphics[height=4cm]{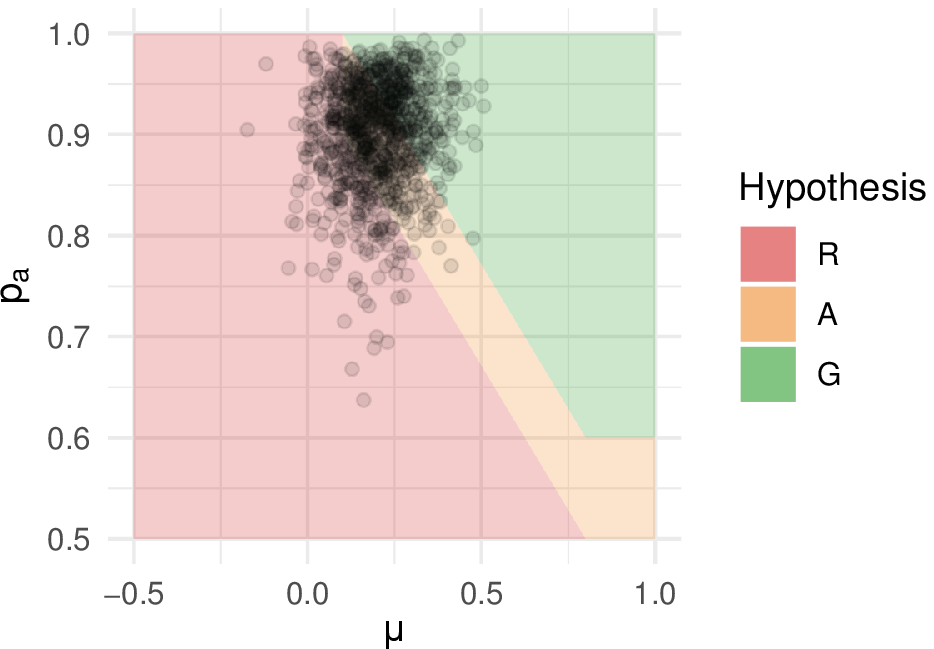}}\\
   \caption{Marginal hypotheses over parameters for (a) follow-up rate $p_{f}$ and mean cluster size $\mu_{c}$; and (b) adherence rate $p_{a}$ and potential efficacy $\mu$. Each point is a sample from the joint prior distribution.}
   \label{fig:hyps}
\end{figure}

The hypotheses are illustrated in Figure~\ref{fig:hyps} (a). Having specified both the hypotheses and the prior distribution for these two parameters, we can obtain prior probabilities of each hypothesis by sampling from the prior and calculating the proportion of these samples falling into the regions $\Phi^i_R, \Phi^i_A$ and $\Phi^i_G$. We have plotted 1000 samples from the prior in Figure~\ref{fig:hyps} (a), falling into hypotheses  $\Phi^i_R, \Phi^i_A$ and $\Phi^i_G$ in proportions 0.354, 0.517, 0.129 respectively. This demonstrates that there is significant prior uncertainty regarding the optimal decision, indicating the potential value of the pilot trial.

\subsubsection{Adherence and potential efficacy}

Having defined priors and hypotheses with respect to cluster size and follow-up, we now consider adherence and potential efficacy. Recall that the number of care homes which successfully adhere to the intervention delivery plan is assumed to be binomially distributed with probability $p_{a}$. We assume that adherence is absolute in the sense that all residents in a care home which does not successfully deliver the intervention will not receive any of the treatment effect. We place a Beta prior on $p_{a}$, with hyper-parameters $\alpha = 28.8$ and $\beta = 3.2$ giving a prior mean of 0.9 and a standard deviation of 0.05.

For the continuous measure of physical activity, we  place priors on the mean effect $\mu$, the intracluster correlation coefficient $\rho$, and the within-cluster variance $\sigma_{W}^{2}$ in the manner suggested in~\cite{Spiegelhalter2001}. Specifically, we choose
\begin{align}
\mu & \sim N(0.2, 0.25^{2}) \\
\sigma_{W}^{2} & \sim \Gamma^{-1}(50, 45) \\
\rho & \sim Beta(1.6, 30.4).
\end{align}
To reflect prior expectation of an ICC around 0.05 but possibly as large as 0.1, the hyperparameters give a prior mean of 0.05 for the ICC with a prior probability of 0.104 that it will exceed 0.1. 

While there is potential for adherence to be improved after the pilot, we assume there will be little opportunity to improve the potential efficacy of the intervention. Moreover, we suppose an absolute improvement in adherence of up to around 0.1 is feasible. To define the hypotheses in this subspace we first set a minimal level of potential efficacy to be 0.1, and decide that we would be happy to make decision $g$ at this point if and only if adherence is perfect. As $p_{a}$ reduces from 1, a corresponding linear increase in potential efficacy is considered to maintain the overall effectiveness of the intervention. The rate of substitution for this trade-off is determined to be approximately 0.57 units of potential efficacy per unit of adherence probability. We consider an absolute lower limit in adherence of $p_{a} = 0.5$, below which we will always consider decision $r$ to be optimal. Taking these considerations together, the marginal hypotheses are defined as

\begin{equation}
  (p_{a}, \mu) \in \begin{cases}
               \Phi^e_R \text{ if } p_{a} < 0.5 \text{ or } 0.96-0.57\mu > p_{a} \\
               \Phi^e_G \text{ if } p_{a} > 0.6 \text{ and } 1.06-0.57\mu < p_{a} \\
               \Phi^e_A \text{ otherwise.}
            \end{cases}
\end{equation}

The hypotheses are illustrated in Figure~\ref{fig:hyps} (b). Again, a sample of size 1000 from the joint marginal prior distribution $p(p_{a}, \mu)$ is also plotted, falling into hypotheses $\Phi^e_R, \Phi^e_A$ and $\Phi^e_G$ in proportions 0.234, 0.470, 0.296 respectively. As before, this indicates substantial prior uncertainty regarding the optimal decision and thus supports the use of a pilot study.

The marginal hypotheses are combined together using equation (\ref{eqn:comb_hyp}). Considering the same 1000 samples from the design prior plotted in Figure~\ref{fig:hyps}, these now fall into the regions $\Phi_R, \Phi_A$ and $\Phi_G$ in proportions 0.507, 0.458, and 0.035 respectively. Note that the prior probabilities of these overall hypotheses are quite different to those of the marginal hypotheses. In particular, there is a considerable increase in the probability that decision $r$ will be optimal, and a considerable decrease that decision $g$ will be. 
\subsection{Evaluation}

\subsubsection{Weakly informative analysis}

We applied the proposed method assuming that a weakly informative joint prior distribution will be used at the analysis stage\footnote{Full details of the weakly informative prior are given in the appendix.}. We took the sample size of the trial to be $k = 6$ clusters per arm. For calculating operating characteristics we generated $N = 10^4$ samples from the joint distribution $p(\theta, x) = p(x | \theta)p_D(\theta)$. We analysed each simulated data set using Stan via the R package rstan~\cite{rstan}, in each case generating 5000 samples in four chains and discarding the first 2500 samples in each to allow for burn-in, leading to $M = 10^4$ posterior samples in total. This gave a maximum Monte Carlo error of approximately 0.005 when estimating a posterior probability $Pr[\phi \in \Phi_I ~|~ x]$, which we considered sufficient. These posterior samples were then used to find the posterior probabilities of each hypothesis, for each simulated data set.

%For each of the $N$ samples $(\theta^{(i)}, x^{(i)})$ we extracted the estimated posterior probabilities $p_R$ and $p_G$, plotted in Figure~\ref{fig:post_probs}. We observe a large mass of points with very low values of $p_G$ and high values of $p_R$. These can be contrasted with the prior probability of each hypothesis, with respect to the design prior $p_D(\theta)$. The majority of simulated data sets lead to a movement from the prior point to a larger value of $p_R$, suggesting the analysis is being conservative.

\begin{figure}
\centering
\includegraphics[scale=0.8]{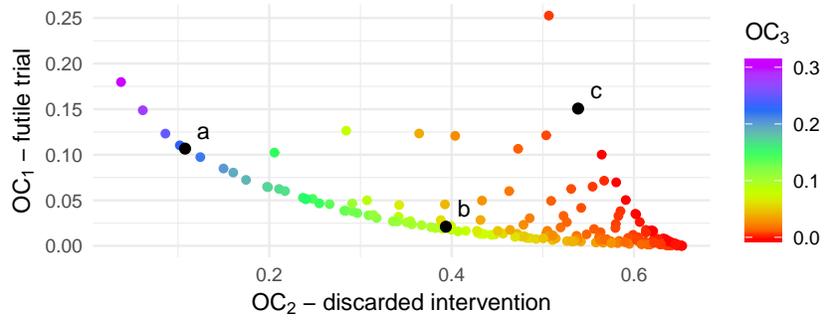}
\caption{Operating characteristics of the example pilot trial for a range of loss parameter vectors, when a weakly informative analysis prior is used.}
\label{fig:p_front}
\end{figure}

We evaluated the operating characteristics for a sample of parameters $(c_1, c_2, c_3)$ as described in Section~\ref{sec:optimisation}. A total of 254 parameter vectors were evaluated, of which 62 led to operating characteristics which were worse in every respect than some other vector (i.e. dominated) and were discarded. The operating characteristics of the non-dominated parameters are shown in Figure~\ref{fig:p_front}. The three operating characteristics are found to be highly correlated. In particular, changing the parameters to give a lower probability of discarding a promising intervention ($OC_2$) tends to lead to a reduction in the probability of making an unnecessary adjustment ($OC_3$). When selecting $(c_1, c_2)$, the key decision appears to be trading off the probability of an infeasible trial, ($OC_{1}$), against $OC_{2}$. There is a very limited opportunity to minimise $OC_{3}$ at the expense of these. For example, compare points $b$ and $c$ in Figure~\ref{fig:p_front}, details of which are given in Table~\ref{tab:costs}. We see that point $c$ reduces $OC_3$ by 0.078 in comparison to point $b$, but only at the expense of increase in $OC_1$ and $OC_2$ of 0.13 and 0.145 respectively.

\begin{table}
\centering
\caption{Estimated operating characteristics (with standard errors) of the REACH trial for the three loss parameter vectors highlighted in Figure~\ref{fig:p_front}, when a weakly informative analysis prior is used. Costs have been rounded to 2 decimal places; operating characteristics and their errors to 3.}
% latex table generated in R 3.4.3 by xtable 1.8-3 package
% Tue Jul 23 10:02:08 2019
\begin{tabular}{lllll}
  \toprule
Label & $(c_1, c_2, c_3)$ & $OC_1$ & $OC_2$ & $OC_3$ \\ 
  \midrule
a & (0.07, 0.9, 0.03) & 0.107 (0.003) & 0.108 (0.003) & 0.232 (0.004) \\ 
  b & (0.18, 0.58, 0.24) & 0.021 (0.001) & 0.394 (0.005) & 0.08 (0.003) \\ 
  c & (0.01, 0.29, 0.7) & 0.151 (0.004) & 0.539 (0.005) & 0.002 (0) \\ 
   \bottomrule
\end{tabular}

\label{tab:costs}
\end{table}

\begin{figure}
\centering
\includegraphics[scale=0.8]{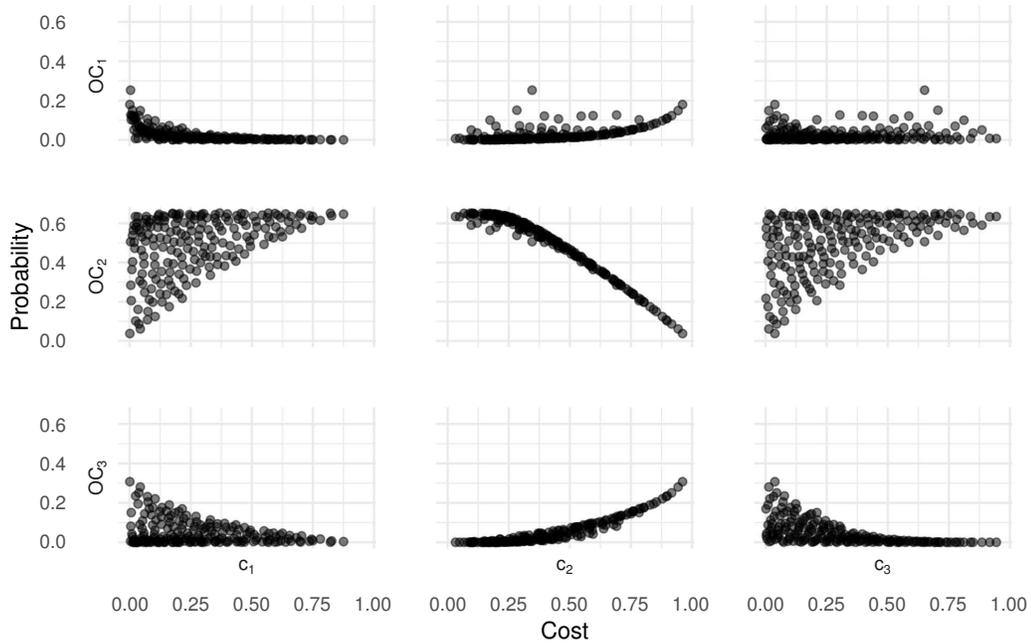}
\caption{Relationships between the three loss parameters ($x$ axes) and resulting operating characteristics ($y$ axes).}
\label{fig:cost_OCs}
\end{figure}

We would expect to see a clear relationship between the value of parameters $c_1, c_2, c_3$ and the operating characteristics they relate to. We explore this in Figure~\ref{fig:cost_OCs} with scatter plots of each  parameter against each operating characteristic. The results show that there is indeed a strong relationship between the loss assigned to discarding a promising intervention, $c_2$, and the probability that this event will occur, $OC_2$ (see centre plot). Moreover, $c_2$ also seems to be the main determinant of operating characteristics $OC_1$ and $OC_3$. The implication is that once the $c_2 \in [0,1]$ has been chosen, the operating characteristics of the trial depend only weakly on the way in which the remaining $1-c_3$ is allocated to $c_1$ and $c_3$. This appears to be due to the fact that, regardless of how errors are weighted, the way we have defined our prior distributions and hypotheses means we are much more likely to make the error of discarding a promising intervention than the other types of error. The cost we assign to this error is therefore more influential on the overall operating characteristics than the other costs.

%RW - what would need to change for this not to be the case?

\begin{figure}
\centering
\includegraphics[scale=0.8]{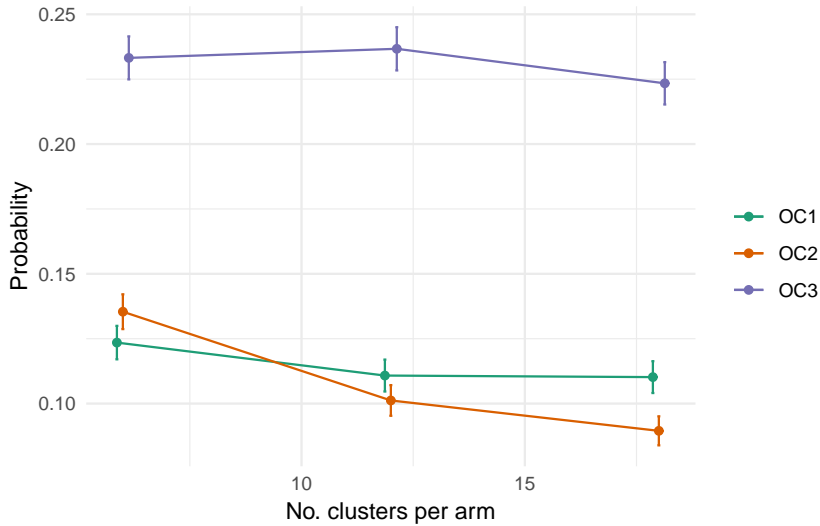}
\caption{Operating characteristics of the REACH trial for per-arm sample sizes $k = 6, 12, 18$ and setting $(c_1, c_2, c_3) = (0.069, 0.116, 0.815)$. Error bars denote 95\% confidence intervals. All points have been adjusted horizontally to avoid overlap.}
\label{fig:k_comp}
\end{figure}

To illustrate the effect of varying sample size in the REACH trial, we set the loss function parameters to that of point $a$ in Figure \ref{fig:p_front} and Table \ref{tab:costs}, $(c_1, c_2, c_3) = (0.07, 0.9, 0.03)$. We then estimated the operating characteristics obtained for $k = 6, 12, 18$ clusters per arm. Note that we considered only three choices of sample size due to the significant computational burden of each evaluation. The results are plotted in Figure \ref{fig:k_comp}. Increasing the sample size appears to have little effect on $OC_1$ and $OC_3$, while leading to a decrease in $OC_2$, the probability of discarding a promising intervention. This behaviour reflects the priorities encoded by the costs parameter, where $c_2 = 0.9$.

\subsubsection{Incorporating subjective priors}

Rather than use weakly or non-informative priors when analysing the pilot data, we may instead want to make use of the (subjective) elicited knowledge of parameter values described in the design prior $p_D(\theta)$. Anticipating criticisms of a fully subjective analysis, we can envisage two particular cases where this might be appropriate. Firstly, using the components of the design prior which describe the nuisance parameters $\psi$ while maintaining weakly informative priors on substantive parameters $\phi$. Secondly, when very little data on a specific substantive parameter is going to be collected in the pilot, using the informative design prior for that parameter could substantially improve operating characteristics.

We replicated the above analysis for these two scenarios. For the second, we used informative priors for all nuisance parameters and for the probability of adherence, $p_a$. Recall that this is informed by a binary indicator at the care home level and only in the intervention arm, and will therefore have very little pilot data bearing on it. For each case we used the same $N$ samples of parameters and pilot data which were used in the weakly informative case, repeating the Bayesian analysis using the appropriate analysis prior and obtaining estimated posterior probabilities $p_R, p_A$ and $p_G$ as before. These were used in conjunction with the same set of loss parameter vectors $\mathcal{C}$ to obtain corresponding operating characteristics.

For brevity we will refer to the three cases as Weakly Informative (WI), Informative Nuisance (IN), and Informative Nuisance and Adherence (INA). Comparing the operating characteristics of cases WI and IN, we found very little difference (further details are provided in the appendix). When we contrast cases WI and INA, however, there is a clear distinction. Using the INA analysis prior will lead to larger probabilities of an infeasible trial ($OC_1$) and of unnecessary adjustment ($OC_2$), while reducing the probability of discarding a promising intervention ($OC_3$), for almost all loss parameters. The expected loss is always lower for the INA analysis than for WI, as we would expect.

%Although the posterior distributions $p(\theta | x)$ are qualitatively different when we examine an individual sample of pilot data, these differences do not translate to the posterior probabilities of hypotheses.

\begin{figure}
\centering
\includegraphics[scale=0.8]{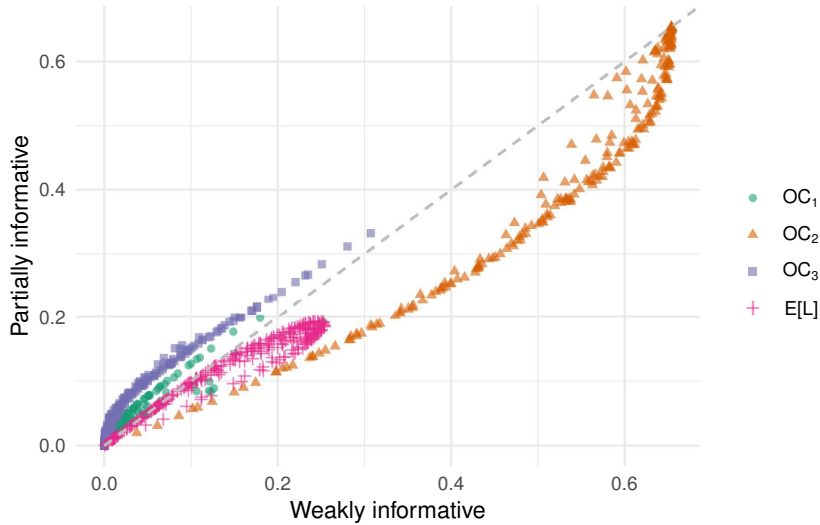}
\caption{Operating characteristics and expected utilities for weakly (WI) and partially informative (INA)}
\label{fig:an_prior_comp}
\end{figure}

\section{Discussion}\label{sec:discussion}

%Using pilot trials to test effectiveness has been repeatedly advised against, as such a test will likely have low power given the small sample size in the pilot~\cite{Lancaster2004, Arain2010, Thabane2010}. Increasing the type I error rate from the conventional values has been suggested as one approach to address this issue~\cite{Lee2014}. This is also suggested in~\cite{Cocks2013} where the authors propose a method which, although described in terms of confidence intervals, is equivalent to powering for a one sided test of effectiveness with a type I error rate of 0.5. In practice, simple rules of thumb such as recruiting 30 participants per arm are commonly used~\cite{Cocks2013, Whitehead2015}.

%In fact, hypothesis testing is generally discouraged due to concerns that the analysis will be under-powered~\cite{Lancaster2004, Arain2010, Thabane2010}.

When deciding if and how a definitive RCT of a complex intervention should be conducted, and basing this decision on an analysis of data from a small pilot trial, there is a risk we will inadvertently make the wrong choice. A Bayesian analysis of pilot data followed by decision making based on a loss function can help ensure this risk is minimised. The expected results of such a pilot can be evaluated through simulation at the design stage, producing operating characteristics which help us understand the potential for the pilot to lead to better decision making. These evaluations can in turn be used to find the loss function which leads to the most desirable operating characteristics, and to inform the choice of sample size.

Our proposal has been motivated by some salient characteristics of complex intervention pilot trials, and offers several potential benefits over standard pilot trial design and analysis techniques. The Bayesian approach to analysis means that complex multi-level models can be used to describe the data, even when the sample size is small. In contrast to the usual application of independent progression criteria for several parameters of interest, we provide a way for preferential relationships between parameters to be articulated and used when making decisions. Using a subjective prior distribution on unknown parameters at the design stage allows both our knowledge and our uncertainty to be fully expressed, meaning we can leverage external  information whilst also avoiding decisions which are highly sensitive to imprecise point estimates.

Our proposed design is related to the literature on assurance calculations for clinical trials \cite{OHagan2005}, applying the idea of using unconditional event probabilities as operating characteristics to the pilot trial setting. In doing so we have shown how assurances can be defined for multiple substantive parameters with trade-offs between them, and with respect to the `traffic light' red/amber/green decision structure commonly found in pilot trials. The multi-objective optimisation framework we have used to inform trial design allows the decision-maker to explicitly consider the different trade-offs between operating characteristics which are available, and select that which best reflects their own preferences. A similar approach has been taken in the context of phase II trials using the statistical concept of admissible designs \cite{Jung2004, Mander2012}. This can be contrasted with the conventional and much criticised approach common in the frequentist context, where arbitrary constraints are placed on type I and II error rates in order to define a single optimal design \cite{Bacchetti2010}.

The benefits brought by the Bayesian approach must be set against the challenges it brings, particularly in terms of computation time and implementation. In terms of the latter, we are required to specify a joint prior distribution over the parameters $\theta$ and a partitioning of the parameter space into the three hypotheses. The specification of the prior distribution may be a challenging and time-consuming task. Although some relevant data relating to similar contexts may be available, for example in systematic reviews or observational studies, expert opinion may still be required to articulate the relevance of such data to the problem at hand. When no data are available, which is not unlikely given the early phase nature of pilot studies, expert opinion will be the only source of information. Although potentially challenging, many examples describing successful practical applications of elicitation for clinical trial design are available~\cite{Walley2015, Crisp2018, Dallow2018}, as are tools for its conduct such as the Sheffield Elicitation Framework (SHELF)~\cite{OHagan2006a}. Dividing the parameter space into three hypotheses may also prove challenging in practice, particularly when trade-offs between more than two parameters are to be elicited. There is a need for methodological research investigating how methods for multi-attribute preference elicitation, such as those set out in~\cite{Keeney1976}, can be applied in this context. 

The computational burden of the proposed method is significant, particularly when the model is too complex to allow a conjugate analysis to be used when sampling from the posterior distribution. We have used a nested Monte Carlo sampling scheme to estimate operating characteristics, as seen elsewhere~\cite{Wang2002, OHagan2005, Sutton2007}. One potential approach to improve efficiency is to use non-parametric regression to predict the expected losses of Equation (\ref{eqn:exp_loss}) based on some simulated data, thus bypassing the need to undertake a full MCMC analysis for each of the $N$ samples in the outer loop. This approach has been shown to be successful in the context of expected value of information calculations~\cite{Strong2014, Strong2015}. The computational difficulties will be particularly pertinent when using our approach to determine sample size, as several evaluations of different sample size choices will be required. If the choice of sample size can be framed as an optimisation problem, methods for efficient global optimisation of computationally expensive functions such as those described in~\cite{Jones2001, Roustant2012} may be useful~\cite{Wilson2015}. Alternatively, one of several rules-of-thumb for choosing pilot sample size \cite{Lancaster2004, Julious2005, Teare2014, Whitehead2015} could be used, with the resulting operating characteristics evaluated using the proposed method.

%The standard approach to the analysis of pilot trials involves pre-specifying some progression criteria for the parameters of interest, and basing decisions (at least in part) on whether or not parameter estimates satisfy these criteria. This framework specifies a decision rule mapping the pilot data to decisions, as we have done in this paper, but does so implicitly and with no statistical analysis of the resulting properties. An alternative approach would be to employ formal hypothesis tests under a frequentist view. This would have the advantage of not requiring a joint prior distribution to be specified when designing the pilot. However, there has been limited reaseach into how preferential relations between parameters can be incorporated into such testing procedures, typically focussing on the case of two parameters~\cite{Conaway1996, Thall2008}. Moreover, testing multiple parameters can lead to multiplicity. In the case of union tests, type I error rate can be inflated; whilst in the case of intersection tests, type II error rate can be inflated~\cite{Senn2007}. Hybrid approaches which assume a freqentist analysis but take a Bayesian view to design by averaging type I~\cite{Chuang-Stein2007} or type II~\cite{Spiegelhalter1986} error rates have been proposed, and may go some way to addressing these issues.

We have defined our procedure in terms of a loss function, where the decision making following the pilot will minimise the expected loss. However, the piecewise constant loss function we have proposed may not adequately represent the preferences of the decision maker. For example, we may object to the loss associated with discarding a promising intervention being independent of exactly how effective the intervention is. An alternative is to try to define a richer representation of the loss function through direct elicitation of the decision makers preferences under uncertainty~\cite{French2000}, leading to a fully decision-theoretic approach to design and analysis~\cite{Lindley1997}. However, as previously noted by others~\cite{Joseph1997a, Bacchetti2008, Whitehead2008}, implementation of these approaches has been limited in practice and this may be indicative of their feasibility.

%We have suggested that the loss function can be parametrised with respect to the operating characteristics it produces. While this is a common strategy when implementing decision-theoretic approaches to trial design~\cite{Lewis2007}, an alternative is to define the loss function through direct elicitation of the decision makers preferences under uncertainty~\cite{French2000}. This leads to a fully decision-theoretic approach to design and analysis~\cite{Lindley1997}. 

%Instead of basing decisions on the expected value of a loss function, a decision rule based directly on posterior probabilities could be used, as in e.g. \cite{Chen2011a, Cellamare2014, Ibrahim2014}. If employing the optimisation approach described in Section~\ref{sec:optimisation} this would make no difference, but if defining the decision rule directly it is possible that this could be easier to do in terms of posterior probabilities than in terms of the cost parameters of the loss function.

% elaborate on the inadequacy of our loss function. Not just that its piecewise constant, but also think about value, utlity, and attitude to risk. Principle of maximising expected utility only applies if our utility function is defined in an uncertain context, i.e. utility reflects preferences between gambles. Perhaps not relevant when we are fitting the loss function to OCs?

% extensions

The proposed method could be extended in several ways. More operating characteristics could be defined and used in design optimisation, more complicated trade-off relationships between multiple parameters could be addressed, or the hypotheses could be expanded to include nuisance parameters which would be used as part of the sample size calculation in the main RCT. A particularly interesting avenue for future research is to consider how  to model post-pilot trial actions in more detail. For example, while we allow for the possibility of making an `amber' decision, indicating that modifications to the intervention or trial design should be made, we do not model what that decision will actually look like and how it should relate to the observed pilot data. Methodology for jointly modelling a pilot and subsequent main RCT in this manner could be informed by developments for designing phase II/III programs in the drug setting \cite{Stallard2012, Wason2013, Goette2015, Kirchner2015}.

\subsection*{Acknowledgements}

We would like to thank Alex Wright-Hughes, Robert Cicero, and the TIGA-CUB and REACH trial teams for discussions which helped shape the scope of this paper.

\subsection*{Data availability statement}

All simulated data used in this manuscript, together with the code used to generate it, is available at \url{https://github.com/DTWilson/Bayesian_pilot}.

\subsection*{Funding}

This work was supported by the Medical Research Council under Grant MR/N015444/1 to D.T.W. and Grant MC\_UU\_00002/6 to J.M.S.W.

\bibliographystyle{unsrt}
\bibliography{C:/Users/meddwilb/Documents/Literature/Databases/DTWrefs}

\end{document}